\documentclass{aa}
\usepackage{graphicx}
\usepackage{txfonts}
\usepackage{lscape}
\usepackage{natbib}
\bibpunct{(}{)}{;}{a}{}{,}

\newcommand{\ctss}{\mbox{counts s}^{-1}}
\newcommand{\fluxA}{\mbox{erg cm}^{-2}\ \mbox{s}^{-1}\ \mbox{\AA}^{-1}}

\begin{document}

\title{The intriguing nature of the high energy gamma ray source 
XSS\,J12270-4859\thanks{Based on observations obtained with 
\emph{XMM-Newton} and 
\emph{INTEGRAL}, ESA science missions with instruments and contributions 
directly funded by  ESA Member States and NASA, with \emph{Fermi} a 
NASA 
mission with contributions from France, Germany, Italy, Japan, Sweden and 
U.S.A. and with the  \emph{REM} 
Telescope INAF at ESO, La 
Silla, Chile}}

\author{D.~de Martino\inst{1}
\and
M.~Falanga\inst{2}
\and
J.-M.~Bonnet-Bidaud\inst{3}
\and
T. Belloni\inst{4}
\and
M.~Mouchet\inst{5}
\and
N.~Masetti\inst{6}
\and
I.~Andruchow\inst{7}
\and
S.A.~Cellone\inst{7}
\and
K.~Mukai\inst{8}
\and
G.~Matt\inst{9}
}


\institute{
INAF - Osservatorio Astronomico di Capodimonte, salita Moiariello 16, I-80131 Napoli, Italy \\
\email{demartino@oacn.inaf.it}
\and
International Space Science Institute (ISSI) Hallerstrasse 6
CH-3012 Bern, Switzerland \\
\email{mfalanga@issibern.ch}
\and
CEA Saclay,  DSM/Irfu/Service d'Astrophysique, F-91191 
Gif-sur-Yvette, France \\
\email{bonnetbidaud]@cea.fr}
\and
INAF-Osservatorio Astronomico di Brera, Via E Bianchi 46. IT 23807 Merate (LC) 
Italy\\
\email{tomaso.belloni@brera.inaf.it}
\and
Laboratoire APC, Universit\'{e} Denis Diderot, 10 rue Alice Domon et L\'{e}onie Duquet, F-75005
Paris, France and LUTH, Observatoire de Paris, Section de Meudon, 5 place Jules Janssen, F-92195
Meudon, France \\
\email{martine.mouchet@obspm.fr}
\and
INAF Istituto Astrofisica Spaziale, Via Gobetti 101, I-40129, Bologna, Italy \\
\email{nicola.masetti@iasfbo.inaf.it}
\and
Facultad de Ciencias Astronomicas y Geofisicas, UNLP, and Instituto de 
Astrofisica La Plata, CONICET/UNLP, Argentina\\
\email{andru@fcaglp.fcaglp.unlp.edu.ar}
\and
 CRESST and X-Ray Astrophysics Laboratory, NASA Goddard Space Flight 
 Center, Greenbelt, MD 20771,
 USA and Department of Physics, University of Maryland, Baltimore County, 
 1000 Hilltop Circle,Baltimore, MD 21250, USA \\
\email{koji.mukai@nasa.gov}
\and
Dipartimento di Fisica, Universit\'a Roma III, Via della Vasca Navale 84,
I-00146, Roma, Italy \\
\email{matt@fis.uniroma3.it}
}

\date{Received December 4, 2009; accepted February 15, 2010}

\abstract {The nature of the hard  X-ray source XSS\,J12270-4859 is still unclear.
It was claimed to be a possible magnetic Cataclysmic Variable of the 
Intermediate Polar type from its optical spectrum and a  possible 
860\,s  X-ray periodicity in \emph{RXTE} data. However, recent 
observations  
do not support the latter variability, leaving this X-ray source still unclassified.} 
{To investigate its nature we present a  
broad-band X-ray and gamma ray study of  this source based on 
a recent \emph{XMM-Newton} observation and archival  
\emph{INTEGRAL} and \emph{RXTE} data. Using the \emph{Fermi}/LAT 1-year 
point source catalogue, we tentatively associate XSS\,J12270-4859   with 
1FGL\,J1227.9-4852, a source of high energy gamma rays with emission up 
to 10\,GeV.  We further 
complement  the study with UV photometry 
from \emph{XMM-Newton} and  ground-based optical and near-IR 
photometry.}
{We have analysed both timing and spectral properties  in the 
gamma rays, X-rays, UV 
and optical/near-IR bands of XSS\,J12270-4859.}
{ The X-ray emission is highly variable showing flares and intensity dips.
The flares consist of 
flare-dip pairs.   Flares are detected in both X-rays and UV range 
whilst the subsequent dips are present only in the X-ray band.
Further aperiodic dipping behaviour is
observed during X-ray quiescence but not in the UV. The broad-band
0.2--100\,keV X-ray/soft gamma ray spectrum is featureless 
 and well described by a power law model with $\Gamma$=1.7. 
The high energy spectrum from 100\,MeV to 10\,GeV  is represented by
a power law index of 2.45. The luminosity ratio between 0.1--100\,GeV 
and 0.2--100\,keV is $\sim$0.8, indicating that the GeV emission is a 
significant component of the total energy output.
Furthermore, the X-ray spectrum does not greatly change during flares,
quiescence and the dips  seen in quiescence.  The X-ray spectrum however 
hardens during the
post-flare dips, where a partial covering absorber is also required to fit 
the spectrum. 
Optical photometry acquired  at different epochs reveals a period of 
4.32\,hr that could be ascribed to the binary  orbital period. 
Near-IR, possibly ellipsoidal, variations are detected. Large amplitude 
variability on shorter (tens mins) timescales are  found to be 
non-periodic.}
{The observed variability at all wavelengths together with  the spectral 
characteristics  
strongly favour a low-mass atypical low-luminosity X-ray binary and are 
against a magnetic  Cataclysmic Variable nature.
The association with a \emph{Fermi}/LAT high energy gamma ray source
further strengths this interpretation.}

\keywords{Stars: binaries: close - Stars: individual: XSS~J12270-4859, 
1FGL\,J1227.9-4852 - gamma rays: stars-  X-rays: binaries - Accretion, 
accretion disks}

\titlerunning{On the nature of  XSS~J12270-4859}

\maketitle

\section{Introduction}

Discovered as a hard X-ray source from the \emph{Rossi XTE} slew 
survey \citep{Sazonov&Revnivstev04}, XSS~J12270-4859 (henceforth XSS\,J1227) 
was also detected as 
an  \emph{INTEGRAL} source and suggested to be a Cataclysmic Variable (CV) 
by  \cite{Masetti06} from its optical spectrum. From follow-up \emph{RXTE} 
observations \cite{Butters08} proposed a magnetic Intermediate 
Polar (IP) type from a possible periodic variability at a 859.6\,s 
period. This periodicity is not confirmed in optical fast 
photometry 
\citep{Pretorius09} and in a \emph{Suzaku} X-ray observation 
\citep{Saitou}.  The latter showed a peculiar X-ray 
variability suggesting  a low-mass X-ray binary (LMXRB).

In the framework of a programme aiming at identifying the nature  
of newly  discovered hard X-ray CV candidates we here present a broad-band  
 gamma ray and X-ray 
analysis complemented with simultaneous UV coverage and new optical 
and near-IR photometry.  
 A search in the recently released  FERMI/LAT 1-year Point 
Source Catalog  provides a possible 
identification that contributes to 
definitively  exclude this 
source 
as a magnetic CV and to favour a LMXRB nature 
with an unusual variable behaviour. 

\begin{figure}[h!]
\centering
\includegraphics[width=\columnwidth,height=8.cm]{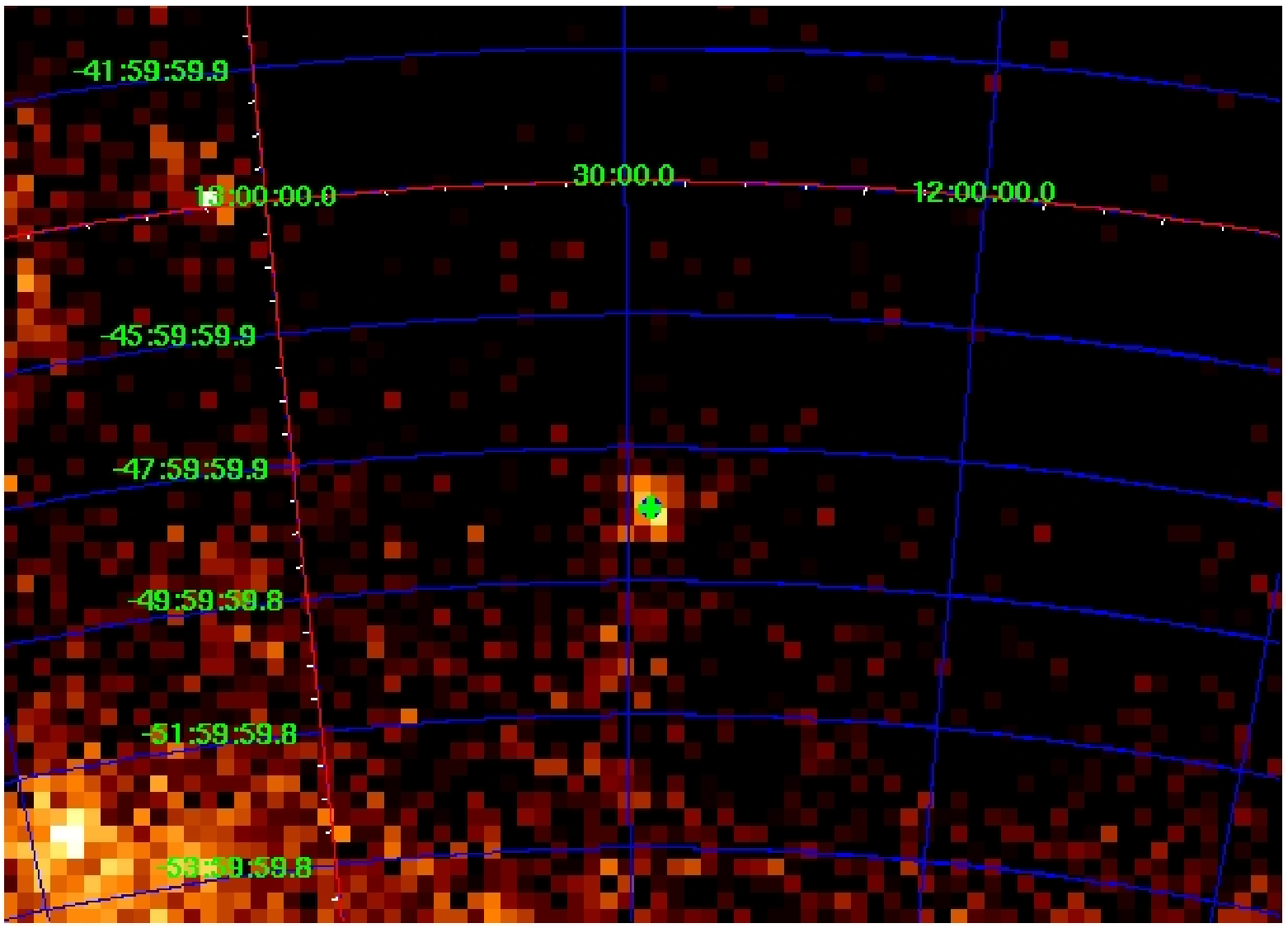}
\includegraphics[width=\columnwidth,height=7.cm,angle=0]{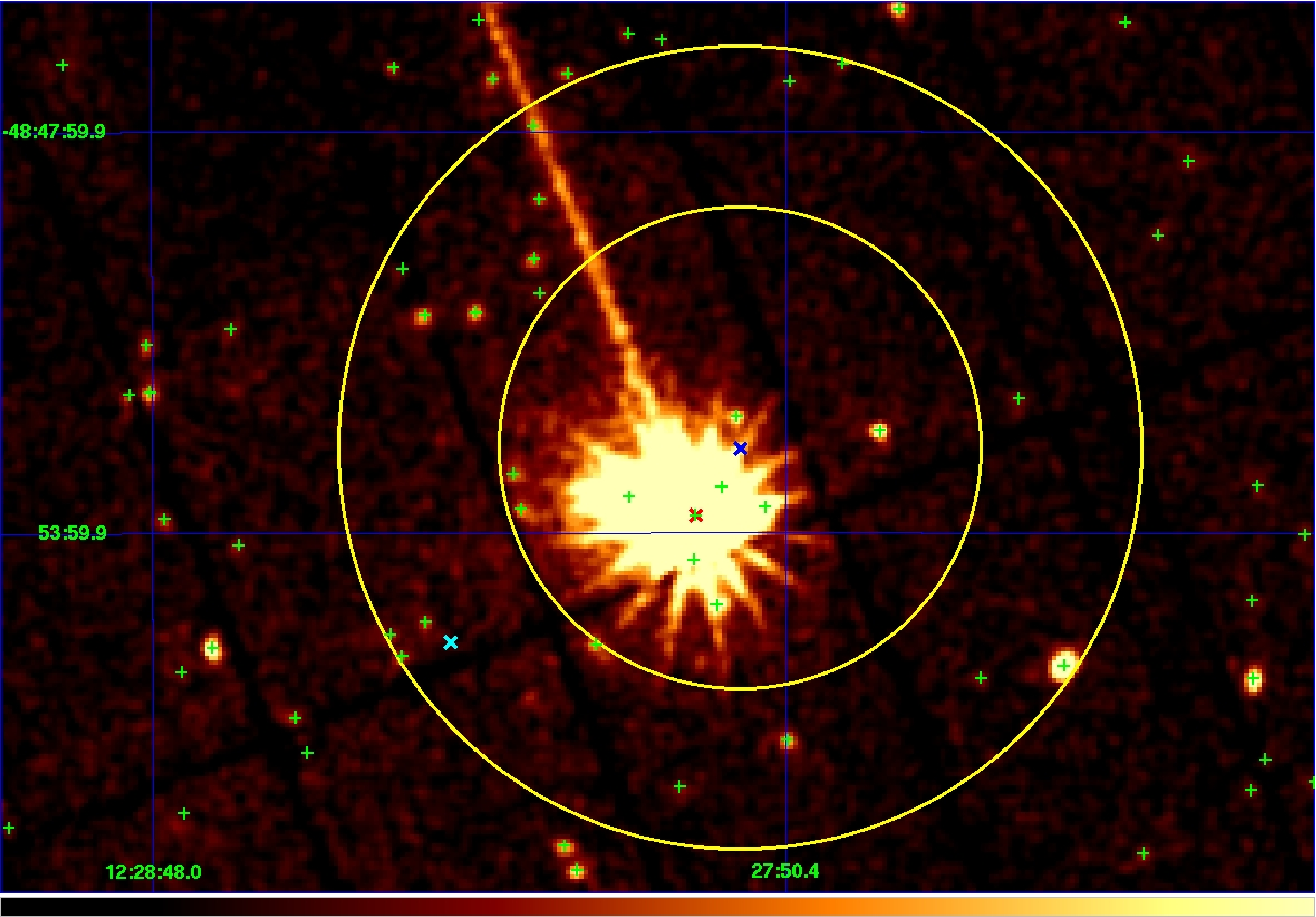}
\caption{{\it Upper panel:}  The counts map in the range 
100\,MeV-300\,GeV of a  12$^o$x12$^o$ region centred at the optical 
position of XSS\,J1227 (marked with a green 
cross) together with the 95$\%$ confidence region (green circle) of the 
\emph{Fermi}/LAT source  1FGL\,J1227.9-4852. {\it Lower panel:}  
The 
combined EPIC pn, MOS1 and MOS2 image in the 0.2--12\,keV range centred on 
XSS\,J1227 (red cross) together with the position (blue 
cross) and 68$\%$ and 95$\%$ confidence 
regions (yellow  circles) of the \emph{Fermi}/LAT source  
1FGL\,J1227.9-4852. Sources 
detected in the EPIC cameras using the standard \emph{XMM-Newton} SSC 
pipeline are also displayed with green crosses. The position of the radio 
source SUMSS\,J122820-485537 is also reported (cyan cross). 
The bright strip is an artifact.}
\label{fig:lat_image}
\end{figure}

\begin{figure*}[th!]
\centering
\includegraphics[width=16.cm,height=15.cm]{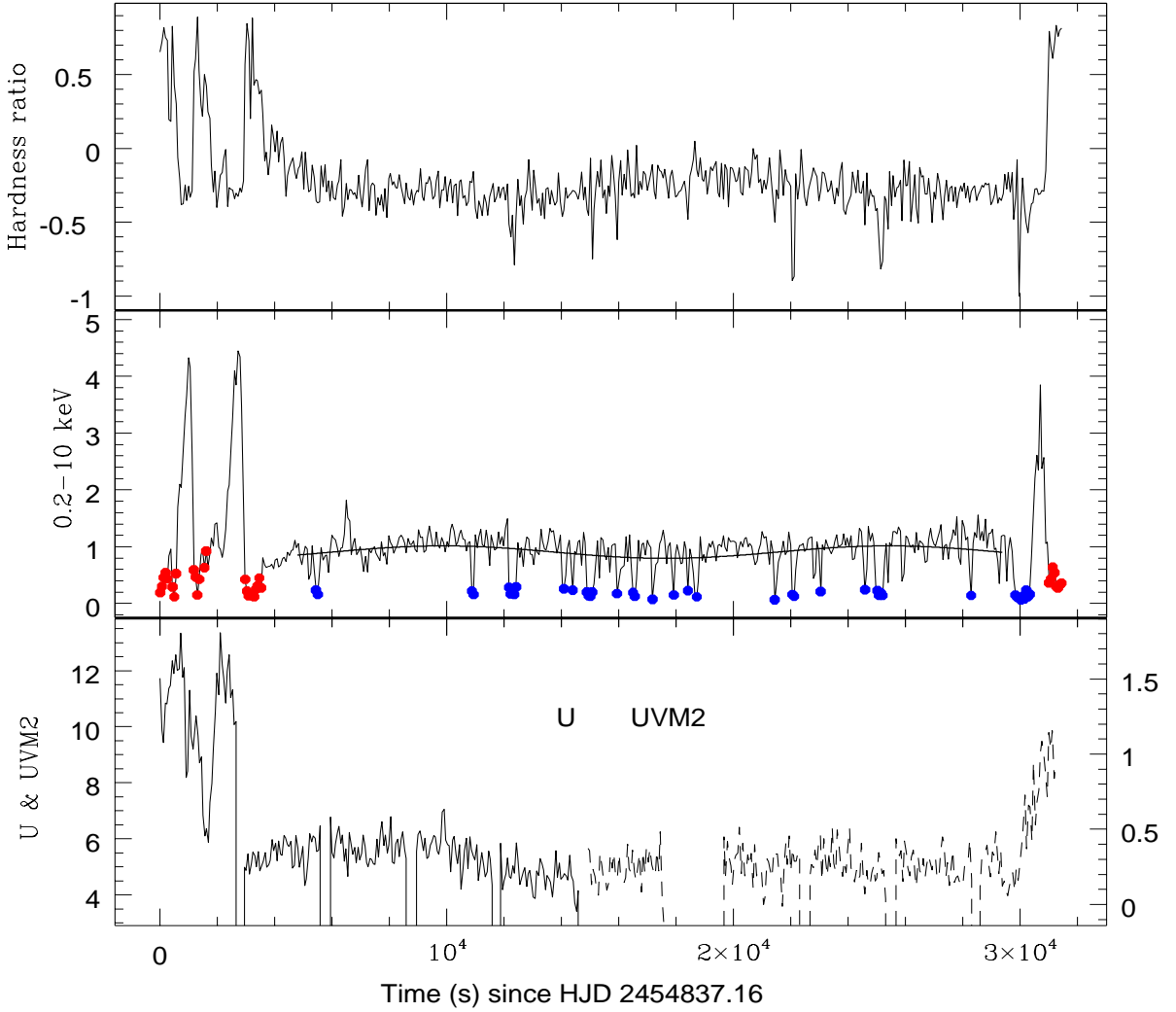}
\caption{{\it Bottom:} 
OM-$U$ (solid line) and OM-$UVM2$ (dashed line) 
background subtracted light curves. 
Ordinates at the  left report the OM-$U$ count rate and at the right 
 the OM-$UVM2$ band count rate. Gaps are due to the OM fast mode windows.  
{\it Middle panel:} EPIC-MOS (MOS1 and MOS2) combined light curve in 
the 0.2-10\,keV band. A sinusoidal function at a period of 244.5\,min is 
also shown. Red points  refer to hard dips whilst blue points refer to 
soft dips that are reported in Fig.\,\ref{fig:hardness_intensity} (see 
also text).
{\it Top:} Hardness ratio defined as [H-S/H+S] between 2--10\,keV and 0.2--2\,keV 
bands from combined EPIC-MOS light curves. A bin size of 60\,s is adopted 
for clarity in the three panels.}
\label{fig:xss_x_uv}
\end{figure*}

\section{ The Fermi GeV source 1FGL\,J1227.9-4852}

 The first point source 
catalogue of high energy gamma ray sources detected by 
the Large Area  Telescope  (LAT) on the Fermi gamma ray Space Telescope 
(Fermi, formerly GLAST) mission was recently released in Jan. 
2010~\footnote{http://fermi.gsfc.nasa.gov/ssc/data/access/lat/1yr$\_$catalog/}. 
This catalogue is based on observations collected  during the 
first 11 months of the science phase of the mission, that began on 
Aug. 4, 2008. It lists and characterizes all sources detected in 
the 100 MeV to 100  GeV range including fluxes in five energy bands as 
well  as best fit power law spectral index.

\noindent Motivated by this first release, we searched into the catalogue 
for 
possible gamma ray counterparts and found that the source 
1FGL\,J1227.9-4852  is at 1.21' from XSS\,J1227 optical position. 
In Fig.\,\ref{fig:lat_image} (upper panel) we show the Fermi/LAT counts 
map  in  the range 100\,MeV-300\,GeV of  a 12$^o$x12$^o$ centred on 
1FGL\,J1227.9-4852 within a maximum zenith angle of 105$^o$. The map, with 
no background subtraction, is constructed from data retrieved 
from the Fermi Science Support Center (FSSC). 
The   semi-major (and equal semi-minor) axis of error ellipse at 68$\%$ 
confidence is 
3.6' and it is 6.0' at 95$\%$ confidence. There is no other
significant detection  within a search radius of 2.5$^o$. 
1FGL\,J1227.9-4852 is detected at a significance of 16.9$\sigma$ 
with a  100MeV--100\,GeV flux of 3.95$\pm 0.44 \times 10^{-11}\,\rm 
erg\,cm^{-2}\,s^{-1}$ and best fit power law index of 2.45$\pm$0.07.   
 1FGL\,J1227.9-4852  is  detected in four out of five LAT 
bands from 100\,MeV up to 10\,GeV, with listed fluxes that are the 
integral 
photon fluxes for the source in the given energy bands computed via the 
maximum likelihood method, that takes into  
account the various backgrounds, contributions from other sources, 
instrument response function and exposure time. 
The optical position is shown together with the 95$\%$ confidence error 
ellipse as reported in the Fermi/LAT 1-year Point source catalogue.
The positional coincidence of both sources is a strong indication 
that XSS\,J1227 could be the counterpart of 1FGL\,J1227.9-4852.

\noindent Variability over a 11-month interval is also provided for
each catalogued source. The 
100\,MeV-100\,GeV light curve of 1FGL\,J1226.9-4852 binned with a time 
interval of one month does not reveal any  variability within 
statistical uncertainty. The
variability index, based on a $\chi^2$-test of deviation of 
the flux in 11 monthly time bins, is 3.8. Hence, the source is 
at a constant flux level on a long term (month) timescale. 
A detailed search for shorter time variations is deferred in a future work.

\section{Observations and data reduction}

Here we report our new observation acquired with \emph{XMM-Newton} as 
well as the publicly 
available \emph{INTEGRAL} data. Archival \emph{RXTE}  
observations were also  retrieved to complement the spectral and timing 
analysis of this source.  We further present optical and near-IR data 
acquired at the \emph{CASLEO} and \emph{REM} telescopes. 

The summary of the observations of XSS\,J1227 is reported in 
Table\,\ref{tab:observ}.

\begin{table*}
\caption{Summary of the observations of XSS\,J1227.}
\label{tab:observ}
\begin{center}
\begin{tabular}{l l l l r c}
\hline\hline 
Telescope	& Instrument	& Date			& UT (start)	& Exposure time 
(s) & Net 
count rate ($\ctss$)\\
\hline
 \emph{XMM-Newton}	& EPIC-pn		& 2009-01-05    & 16:09			
& 30\,002		& $2.86 \pm 0.01$ \\
 		& EPIC-MOS		&				& 15:46			
& 31\,504		& $0.97 \pm 0.04$ \\
		& OM-U 			&				& 15:56	
	& 13\,298		& $6.35 \pm 0.04$	\\
		& OM-UVM2		&				& 
20:05			& 13\,296		& $0.36 \pm 0.01$	\\
& & & & & \\

\emph{INTEGRAL} 		& IBIS/ISGRI	& 				
&				& $\sim 750\,000$	& $0.14\pm 0.05 $ \\ 
& & & & & \\
\emph{RXTE}                & PCA  & 2007-11-28 & 16:13 & 6\,784 & 1.3$\pm$0.1\\
                           &      & 2007-11-28 & 22:02 & 3\,616 & 2.6$\pm$0.1 \\
                           &      & 2007-11-29 & 02:04 & 1\,168 & 1.4$\pm$0.2\\
                           &      & 2007-11-29 & 04:35 & 22\,832 & 1.43$\pm$0.05\\
                           &      & 2007-11-29 & 12:06 & 14\,416 & 1.28$\pm$0.07\\  
\hline

\hline\hline 
Telescope	& Instrument	& Date			& UT (start)	& Exposure time 
(s) & Number exposures\\
\hline

 & & & & & \\

\emph{REM}             & ROSS-V		& 2009-03-18	& 05:46	&  60 &  93 \\
                       & ROSS-V		& 2009-03-19	& 05:54 &  60 & 175 \\
                       & ROSS-V		& 2009-03-20	& 05:54 &  60 & 185 \\
		& REMIR-J		& 2009-03-18	& 05:47 &  75 &  30  \\
		& REMIR-J		& 2009-03-19	& 05:54 &  75 &  75\\
		& REMIR-J		& 2009-03-20	& 05:54 &  75 &  83 \\
 & & & & & \\
\emph{CASLEO}          &  B                    & 2008-07-07    & 00:28 &  30 & 158\\
\hline
\end{tabular}
\end{center}
$^{*}$ Net exposure times except for the \emph{INTEGRAL} 
observations. 
\end{table*}

\subsection{The \emph{XMM-Newton} observations}

The \emph{XMM-Newton} observation (OBSID:  0551430401) was carried out on Jan.5, 2009 with the 
EPIC cameras (pn: \cite{struder01} and MOS: \cite{turner01}) operated in 
imaging full window 
mode using the thin filters and with the OM \citep{mason01} operated in fast window mode using 
sequentially the U (3000--3800 \AA) and UVM2 (2000--2800 \AA)  filters for 
$\sim$13.3\,ks each. 

The data were processed 
using the standard reduction 
pipelines and analyzed with the SAS  8.0 package using the latest 
calibration files.
We used a  34''  (37'') aperture radius to extract EPIC-pn 
(EPIC-MOS) light curves and 
spectra from a circular region
centered on the source and from a background region located on the same CCD chip.
In order to improve the S/N ratio, 
we filtered the data by selecting pattern pixel events up to 
double with zero quality
 flag for the EPIC-pn data,  and up to quadruple pixel events for the EPIC-MOS data. 
The average  background level of the EPIC cameras was low for all the duration of the 
observation, with the exception of a flaring activity, lasting $\sim 4800$ s and  occurring 
at the end of the EPIC-pn exposure. This flare does not
significantly affect the light curves, but we conservatively 
exclude that period in the extraction of the spectra.

Considering the possible association of XSS\,J1227 with the high 
energy gamma ray source 1FGL\,J1227.9-4852, we also inspected the  
source catalogue detected in the EPIC cameras, produced by the 
standard \emph{XMM-Newton} 
Survey Science Center (SSC) pipeline. In Fig.\,\ref{fig:lat_image} (lower  
panel), the combined EPIC pn, MOS1 and MOS2 image in the 
0.2--12\,keV
centred on XSS\,J1227 is shown together with the \emph{Fermi} 68$\%$ and 
$95\%$ confidence regions. Although there are a number of X-ray 
sources within these regions, these are much fainter than XSS\,J1227. Most 
are found at  a count rate (EPIC pn) much lower than 0.04 $\rm 
cts\,s^{-1}$,  with 
only one source at 0.26\,$\rm cts\,s^{-1}$ (corresponding to 
a 0.2-12\,keV flux of 8.6$\times 10^{-13}\,\rm 
erg\,cm^{-2}\,s^{-1}$), located at 1.0' north-east from XSS\,J1227. 
We therefore conclude that there are no other favoured X-ray counterparts 
than XSS\,J1227.

 Background subtracted OM-$U$ and OM-$UVM2$ light curves were obtained with a binning 
time of 10 s.
 The average count rates were $6.35\ \ctss$ in the $U$ band and $0.36\ 
\ctss$ in the $UVM2$ band, 
corresponding to instrumental magnitudes
$U = 16.5$ and $UVM2 = 17.3$
and to average fluxes  $8.2 \times 10^{-16}\ \fluxA$ and $5.3 \times 
10^{-16}\ \fluxA$, respectively. 

Heliocentric corrections were applied to the EPIC and OM arrival times.

\subsection{The \emph{INTEGRAL} observations}
 
The \emph{INTEGRAL} IBIS/ISGRI \citep{ubertini03,lebrun03} hard X-ray data of the source 
were extracted from all pointings within 12\degr\ from the source positions, spanning from March 2003 
to October 2007. The total effective exposure times is $\sim 750$\,ks  
(651 pointings).
To study the weak persistent X-ray emission, the time averaged 
ISGRI spectrum was obtained from mosaic images in five energy bands, 
logarithmically spaced 
between 20 and 100 keV.
Data were reduced with the standard OSA software version 7.0 and, then, analyzed using the algorithms 
described by \citet{goldwurm_etal03}.

\subsection{The \emph{RXTE} observations}

Archival \emph{RXTE} \citep{bradt_etal93} observations 
acquired in November 
2007 and published in \cite{Butters08} were retrieved to search for the 
peculiar behaviour detected in our \emph{XMM-Newton} observation, about one 
year later.  The  
\emph{RXTE}/PCA exposures amount to a total effective time of 48.8\,ks 
(see Table\,\ref{tab:observ}. 
This  is remarkably longer than that reported in \cite{Butters08}. To 
allow a direct comparison, the \emph{RXTE} data reduction and analysis was 
performed with a procedure similar to that described in \cite{Butters08}. 

\subsection{The optical and near-IR photometry}

XSS\,J1227 was observed for one night on July 7, 2008 at the 2.15\,m telescope at the 
Complejo 
Astronomico el Leoncito, \emph{CASLEO}, in Argentina, equipped with a direct CCD camera. B band 
time-series photometry was 
acquired for 2.64\,h adopting 30\,s exposure times for each image. 
Data reduction was performed using standard \emph{iraf} routines. The B band photometry was 
however affected by spurious light variable with time of unknown nature. 
We therefore did not apply dark and sky corrections. Aperture 
photometry was  obtained for the star and several (ten) comparison stars 
in the field.  Hence, differential photometry was  obtained by dividing 
the target count rates with the average count rate of the comparison 
stars.  The error associated to each data point was less than 0.05\,mag

XSS\,J1227 was further observed on  March, 18,19 and 20, 2009 with the 0.5\,min
Rapid Eye Mount (\emph{REM}) robotic telescope 
at the ESO, La Silla observatory in Chile \citep{Zerbi04}, equipped with the ROSS 
\citep{Tosti04} and REMIR \citep{Conconi04} cameras that 
simultaneously covered the V and J 
photometric bands. Exposure times of individual images were 60\,s in the V band. The J band 
photometry was carried out using a dithering of five images on the source, 
each of 15\,s exposure. A sky J band image was also sequentially acquired for each set of 
dithered  images.  The source was observed for 1.8\,h, 3.4\,h and 3.7\,h  during the 
three consecutive nights. 
Data reduction was also performed using \emph{iraf} routines including flat field 
and dark corrections for the V band images. The J band images are routinely pre-reduced by 
the \emph{REMIR} pipeline that provides  sky subtracted and de-biased images.
Aperture photometry was carried out using \emph{daophot} routine. Relative photometry was
obtained by using several comparison stars in the field. 
Flux ratios were obtained by dividing the 
target counts by the combination (weighted mean) of five reference stars.
Errors on each data points are $\sim$0.09\,mag and 0.3\,mag in the V 
and J band respectively. In the latter band  the source was faint and 
not always detected  providing a badly sampled light curve.
 Average V band magnitudes were: 16.70, 15.85 and 17.31\,mag and the 
average J band magnitudes were 
16.14, 15.75 and 16.53\,mag during the three consecutive days. These were 
derived by  comparing the photometry of the reference star HD 108433 in 
the target field. 
A comparison with the optical spectrum acquired by \cite{Masetti06}  indicates that 
XSS\,J1227 was
about at the same optical level. Furthermore from a comparison with J-band 2MASS 
magnitude (J=15.73\,mag), 
the source is found at a comparable level. 

The flux ratios of individual nights were converted to 
fractional  intensities dividing them by the corresponding nightly mean.
Heliocentric corrections were also applied to all photometric data sets.

\section{Results}

\subsection{The X-ray variability of XSS\,J1227}

 The EPIC-pn and MOS light curves were extracted in the energy range 
0.2--12.0 keV and 
binned in 20\,s time intervals. In all instruments we observe similar temporal 
behaviour. Hence, due  to the slightly shorter exposure of the  EPIC-pn 
camera we 
discuss the variability observed in the EPIC-MOS cameras. 

\begin{figure}
\centering
\resizebox{\hsize}{!}{\includegraphics{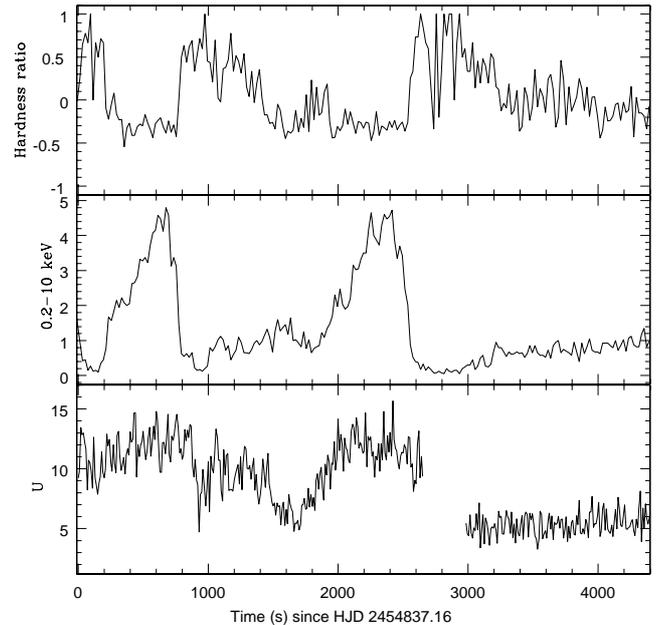}}
\caption{Enlargements of the activity phase at the beginning of the 
{\it XMM-Newton} observation in the U band (bottom), the X-ray 
0.2--10\,keV  band together with hardness ratios  defined in the text 
(top). Bin size is 20\,s.
}
\label{fig:xss_flare}
\end{figure}

The X-ray light curve shows two 
strong events at the beginning of the exposure where 
the count rate reaches a peak intensity $\sim$4.6 times the persistent 
level. The second occurs  $\sim$30\,min after the first one. A third event, 
covered only by  the EPIC-MOS cameras, is also 
observed at the end of the exposure, $\sim$7.3\,hr after the second one.  
We also detect similar events in the 
$OM-U$ and $OM-UVM2$ bands (Fig.\,\ref{fig:xss_x_uv}), which therefore are 
regarded as real. We will refer to them 
as "flares".   These  are remarkably similar to each other in temporal 
evolution and intensity. 
The rise is  less steep than the decay and structured, consisting of a sequence 
of peaks before reaching the maximum intensity (see enlargement in 
Fig.\,\ref{fig:xss_flare}). 
The third flare is even more 
structured. The decay to quiescent level is also not smooth, with a peak 
occurring $\sim$1\,min later  than the maximum 
(best seen in the first two flares) and lasting  $\sim$2.3\,min, followed by a 
decrease of the count rate below the quiescent persistent level ("dip").  
The duration of the flares are 11.5\,min and 12\,min and 9.2\,min,  
respectively. The third flare is also preceeded by a rather broad dip.
An additional weak event is observed $\sim$1.8\,hr 
after the beginning of the MOS exposure with  peak intensity 2\,times the 
quiescent flux and duration of 3.9\,min.  Also for this event a weak and 
short dip is detected that however does not reach the low count rate level 
observed in the others.

We here define the hardness ratio as: HR = [H--S/H+S],  where H and 
S are  the count  rates in the  2--10\,keV and 0.2--2\,keV bands, 
 respectively.
Their temporal behaviour,  shown in Fig.\,\ref{fig:xss_x_uv}, demonstrates
that  the flares are not characterized by strong changes in the spectral 
shape, contrary to the dips associated to them, that  show instead a 
hardening of the spectrum. Also,
 pronounced non-periodic dips characterise the quiescent persistent 
emission. The count rate changes by a factor of 
$\sim$3  in the strongest dips. 
Their occurrence intensifies between 
2.9\,hr and 7.9\,hr after the beginning of the MOS observation. The dips 
have variable length ranging from 2\,min to 5\,min for the more pronounced dips. 
Hardness ratios (Fig.\,\ref{fig:xss_x_uv}) do not greatly change, 
except in a few  intense dips where a softening is observed.
Hence, the dips observed in quiescence and those associated to the 
flares have different origin.

\begin{figure}[h!]
\centering
\resizebox{\hsize}{!}{\includegraphics{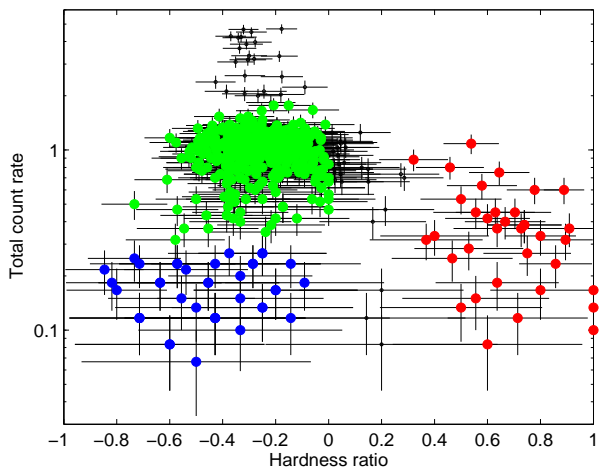}}
\caption{Diagram of total MOS intensity versus hardness ratio  in the 
0.2--2\,keV and 2--10\,keV bands. Green points represent quiescence (not reported in 
Fig.\,\ref{fig:xss_x_uv} for clarity), red points 
are the dips observed after the flares and the blue points represent the dips 
observed during quiescence. Flares are denoted with black points. 
} 
\label{fig:hardness_intensity}
\end{figure}

This peculiar  behaviour is best seen in  
the intensity versus hardness ratio (HR) diagram as depicted in 
Fig.\,\ref{fig:hardness_intensity}. Different locii can be 
identified: hard dips (shown in red) with hardness ratio HR$>$0.3; soft dips 
(shown in blue) with HR$<$0 and count  rate $<$0.3 cts\,s$^{-1}$; 
quiescence with  count rate $>$0.3cts\,s$^{-1}$ but 
$<$2\,cts\,s$^{-1}$ and -0.7$<$HR $<0.3$ (shown in green) and the flares 
with count  rate $>$2\,cts\,s$^{-1}$ and -0.4$<$HR$<$0. 
The diagram reveals similar spectral shape between quiescence and 
flares. The spectrum instead is harder during the dips after the flares but not 
during those occurring in quiescence. The red and blue colours refer to the 
points also  reported in Fig.\,\ref{fig:xss_x_uv}. Since the 
quiescence prior the first flare is not 
observed and given that the dip preceding it shows similar 
spectral behaviour 
as the post-flare dips, it is plausible  that this first flare is also 
preceeded by  another event. If it is the case, it might be concluded that 
 only the post-flare dips show spectral hardening 
with respect to all other temporal features observed in the 
X-ray light curve.

We inspected flare fluence (total flare counts), duration and 
total flux in the dips associated to them. The ratio of flare fluence and 
count rate deficiency is $\ga$8 in the first 
event and,  remarkably, $\sim$7.5 in the second and third. Note 
that the dip associated to the first flare is less 
evident and hence more difficult to isolate; it lasts less than the 
others, possibly because of the  superposition or closeness of the  second 
flare. The length of the dips increases with the duration of the 
flare, the second flare and associated dip being the longest. This 
suggests that  flares and dips are correlated and hence the events consist 
of flare-dip pairs. 

\begin{figure}[h!]
\centering
\resizebox{\hsize}{!}{\includegraphics{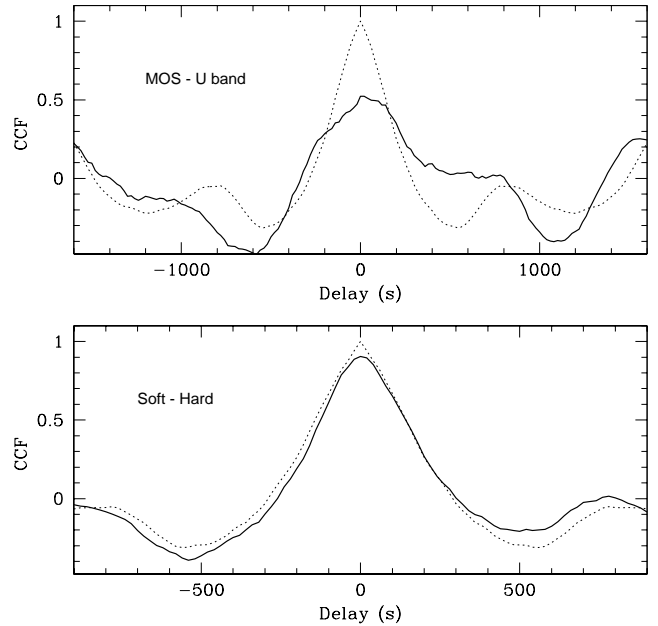}}
\caption{{\it Bottom:} The CCF between the hard (2--10\,keV) and soft 
(0.2--2\,keV) light curves 
during the first two flares  (thick line)  together with the 
auto-correlation function of the soft band light curve, taken as reference 
light curve (dotted line). {\it Top:} The 
CCF between the U band and 0.2--10\,keV MOS light curves (thick line) 
together with the auto-correlation function (dotted line) of the MOS light 
curve. } 
\label{fig:cross_corr}
\end{figure}

Variations  in  the soft and hard X-ray bands
during flares  were  also inspected by computing the 
cross-correlation 
function 
(CCF) between the soft, 0.2--2\,keV and hard 2--10\,keV bands. This is 
shown in  Fig.\,\ref{fig:cross_corr} (bottom panel) for the portion of the 
light curve that  includes the first two flares. The 
CCF peaks at zero lag, but is slightly asymmetric towards positive lags. 
Taking as reference the soft band, the CCF suggests that the hard X-ray 
variations on timescales $\ga$300\,s lag the soft ones. This can be 
also seen from the large amplitude variations of HRs, shown in 
Fig.\,\ref{fig:xss_flare}, during and immediately after the dips.

We also detect a weak (4$\%$) long-term quasi-sinusoidal trend in the
quiescent flux.
The length of the MOS  exposure however does not allow us to determine the 
period as we obtain  
similar $\chi^2$ fitting a sinusoid with periods between 
200\,--280\,min. To 
illustrate this we also show in Fig.\,\ref{fig:xss_x_uv} a sinusoidal 
function at the fixed period of 4.32\,h found in the optical photometry 
(see Sect.\,3.3). 

\begin{figure}[h!]
\centering
\resizebox{\hsize}{!}{\includegraphics{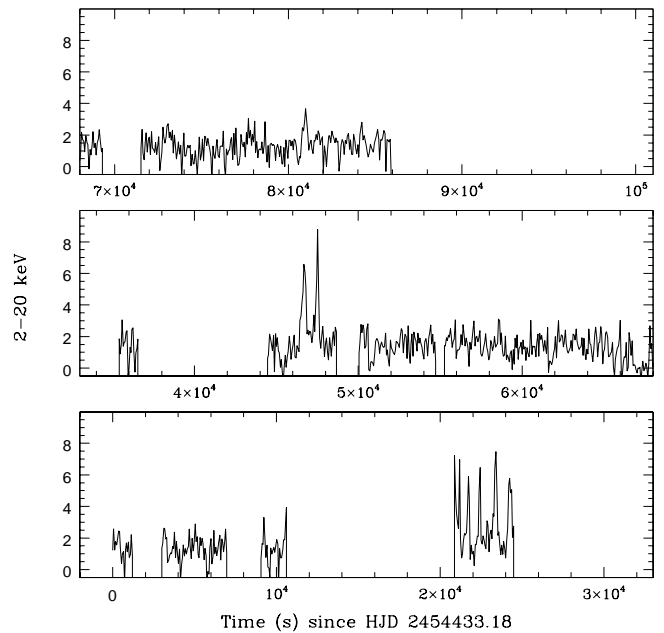}}
\caption{ The \emph{RXTE} PCA light curve in the 2-10\,keV band 
as observed in 2007.
A bin size of 60\,s and a similar temporal scale as the \emph{XMM-Newton} 
EPIC light curve are adopted for clarity.}
\label{fig:RXTElight}
\end{figure}

This peculiar X-ray variability led us to compare the \emph{XMM-Newton} 
light curve with that observed with \emph{RXTE} in Nov. 2007, as  
\cite{Butters08} neither show it nor report any atypical behaviour.
The \emph{RXTE} PCA light curve in the 2--20\,keV band is displayed in 
Fig.\,\ref{fig:RXTElight} with similar sampling and temporal scale as the
EPIC MOS one. Also in these data flares and dips are observed with
similar timescales as in our data. In particular a series of flares 
are observed $\sim$20\,ks after the start of the \emph{RXTE} pointing, each 
of 
them lasting $\sim$8--12\,min. The duration of this active period cannot be 
determined due to the gaps in the data. It is however clear that a 
relatively long ($\ga$5.3\,hr) quiescent period follows the active phase, 
during which dips, lasting $\sim$6-13\,min, are detected. 
This quiescent period
has been Fourier analysed in order to detect the  859.6\,s 
periodicity reported by \cite{Butters08} but we did not find any significant 
peak. Furthermore and as apparent from Fig.\,\ref{fig:RXTElight}, the data 
are too noisy to detect low frequency variations. 

We therefore conclude that XSS\,J1227 is a persistent X-ray source 
with flaring and dipping characteristics. This is further 
corroborated  by a \emph{Suzaku} observation carried out in August 2008, 
recently reported by \cite{Saitou}.

\begin{figure*}[ht!]
\centering
\includegraphics[width=\columnwidth,height=10.cm]{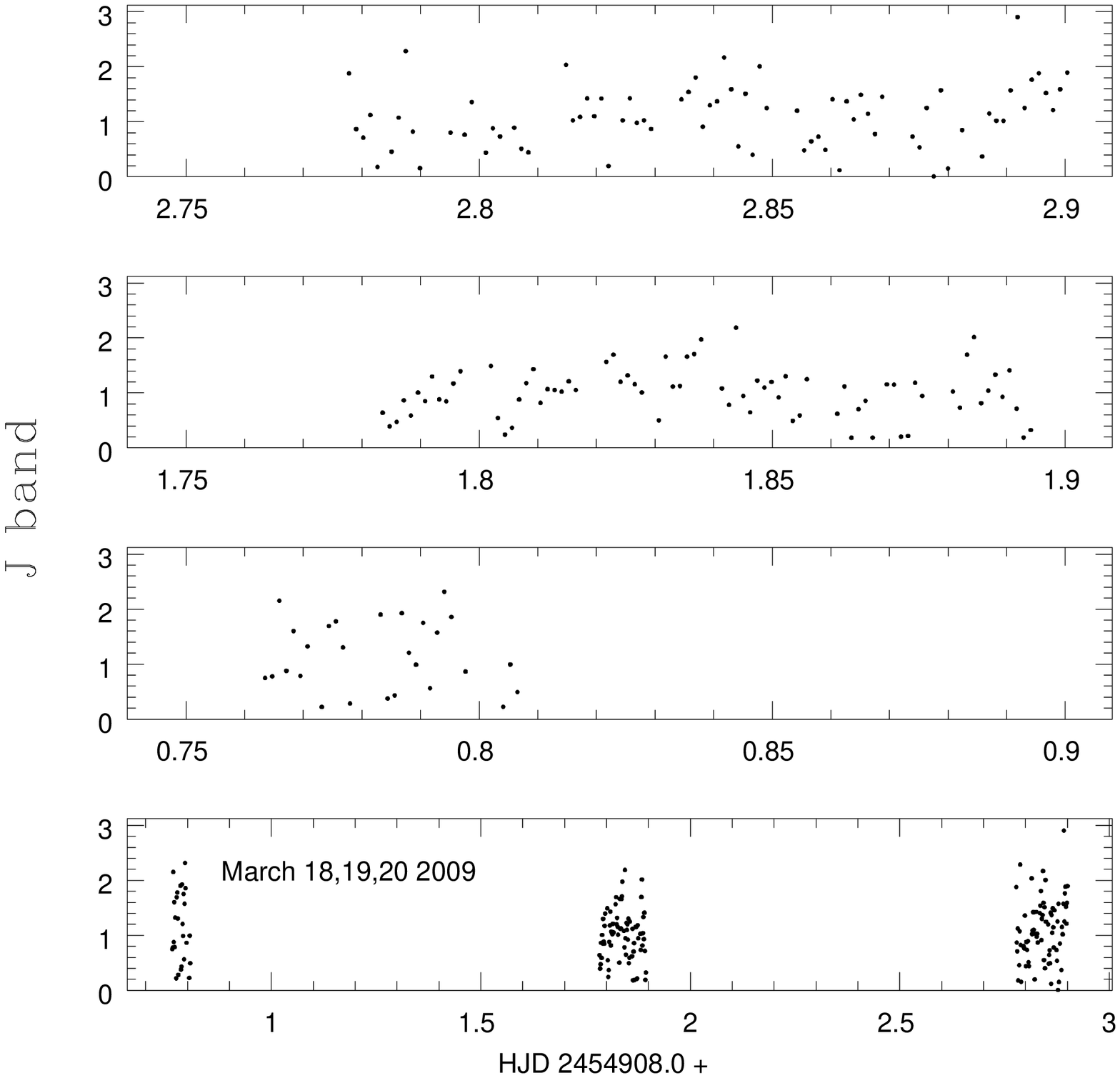}
\includegraphics[width=\columnwidth,height=10.cm]{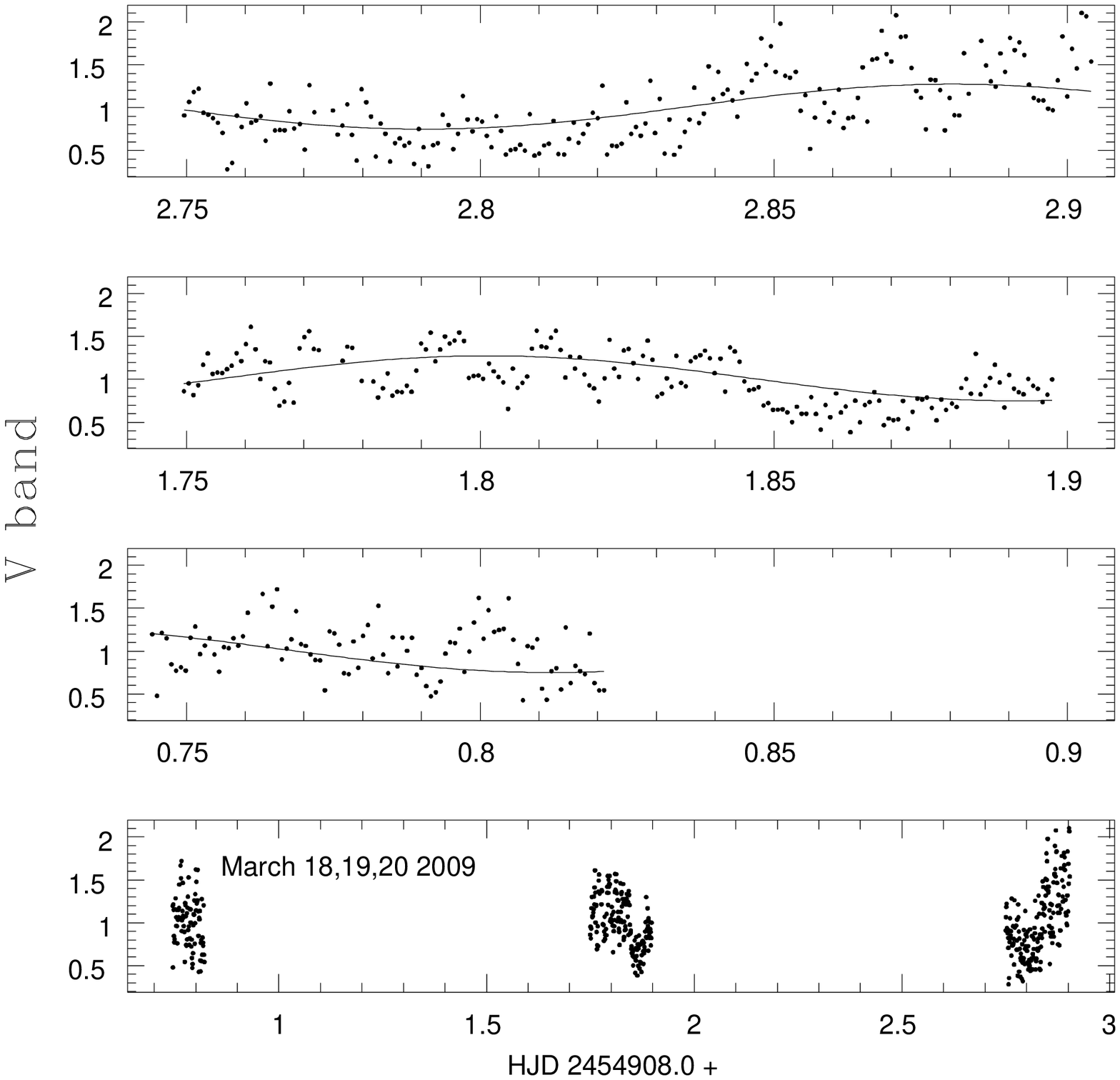}
\includegraphics[width=\columnwidth,height=3.5cm]{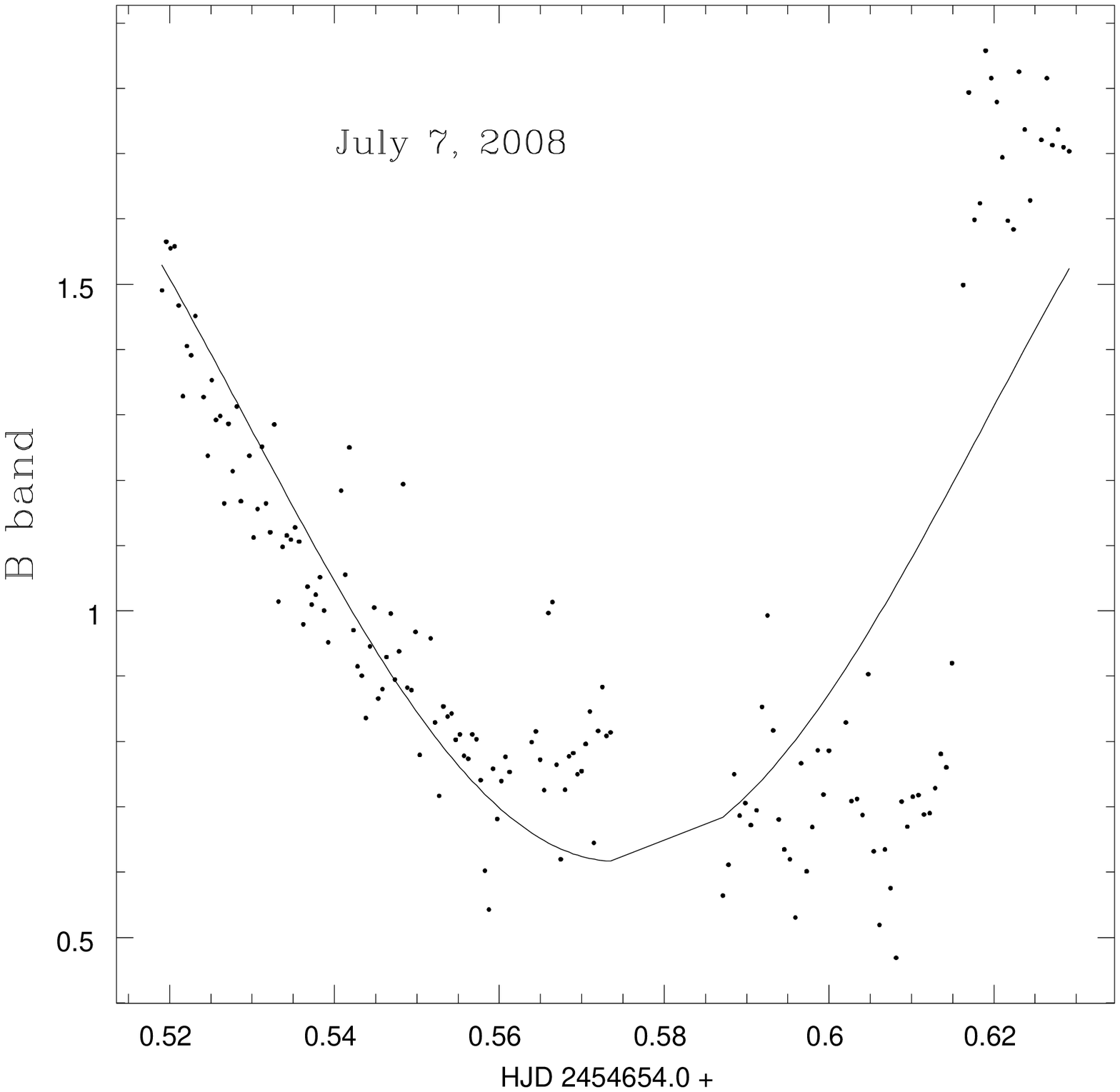}
\caption{{\it Bottom:} The B band light curve observed in 2008. {\it Top 
right panels:}
The V band photometry   during the three consecutive nights in March 
2009.  {\it Top left panels:}
The near-IR light curves during the same nights in 2009. The optical light curves are also shown
together with a sinusoidal function at a period of 259\,min (see text for details).}
\label{fig:opticalcurves}
\end{figure*}

\subsection{The UV variability of XSS\,J1227}

The flares detected in the ultraviolet, U and UVM2 bands, have also  
similar peak 
intensities reaching $\sim$3\,times the quiescent persistent level 
(Fig.\,\ref{fig:xss_x_uv}). 
They however appear to last longer than the X-ray flares, though 
we  lack coverage of the 
quiescent level before onset of the first UV flare and of the decay to 
quiescence of the second flare. Also, the U band light curve is 
highly structured between the two events, whilst it is not so in the 
X-rays.
We find that the decay to quiescence of the first 
flare in the U band occurs 1.2\,min later than the X-rays  and the rise 
of the second  flare occurs 1.8\,min earlier than the X-rays (see
enlargement in Fig.\,\ref{fig:xss_flare}). Similarly,  the third flare 
observed  at the end  of the {\em XMM-Newton} observation starts earlier 
in the UVM2 band (Fig.\,\ref{fig:xss_x_uv}) and, although only the rise is 
covered, it also  seems to last longer than the X-ray flares. 
Worth noticing is that the weak X-ray flare after  
the first two larger ones, is not detected in the ultraviolet (U band).

We also computed the CCF between the U and X-ray, 
0.2--10\,keV, light curves covering the first two flares taking as 
reference the X-ray light curve (see 
Fig.\,\ref{fig:cross_corr} upper panel). 
Also in this case the CCF peaks  at zero lag but it is strongly 
asymmetric towards  positive lags, with major  differences at lags $\ga$ 
300\,s up to 900\,s. Hence the UV  variations  are 
delayed with respect  to the X-rays and this could be understood with a 
longer duration of the UV flares.

The U band magnitude at flare peak is $\rm U_{peak}\sim$15.3\,mag and 
at quiescence  $\rm U_{quie.}\sim$16.5\,mag. In the UVM2 band, we observe
the third flare at $\rm UVM2_{peak}\sim$15.3\,mag 
while at quiescence $\rm UVM2_{quie.}\sim$17.3\,mag. The magnitude
differences are then 1.2$\pm$0.1\,mag and 2.0$\pm$0.3\,mag in the 
U and UVM2 bands, respectively. Assuming that all flares in each band 
reach  similar intensity, then $\rm (UVM2 - U)_{peak}\sim 0.0$ and  $\rm 
(UVM2 - U)_{quie.}\sim 0.8$, implying that flares are blue. 


Furthermore, the X-ray dips occurring in quiescence do not have a 
counterpart in 
the UV range. This  suggests that the dipping behaviour is related to the 
 regions where the X-rays are emitted. The quiescent UV flux seems instead 
to be variable
on timescales of hours resembling the weak quasi-sinusoidal variation
seen in the X-rays, but we are unable to determine any periodicity due to 
the short coverages in the two bands.

\begin{figure}
\centering
\resizebox{\hsize}{!}{\includegraphics{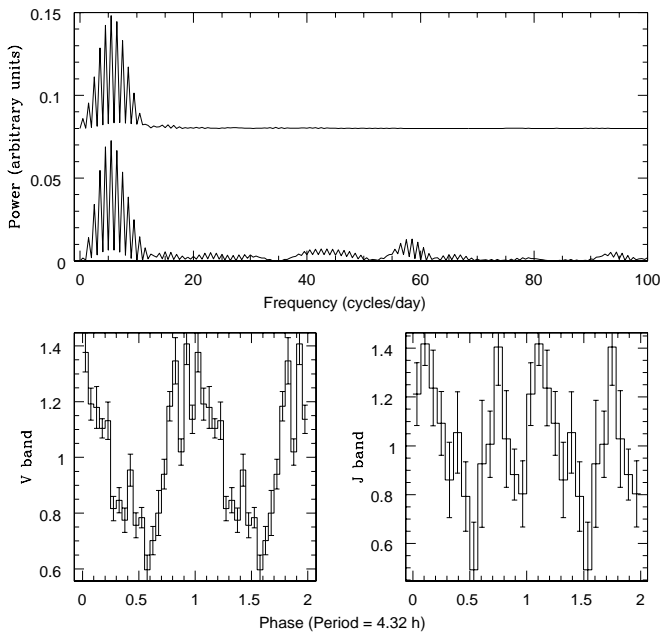}}
\caption{ {\it Top} The Fourier spectrum of the V band photometry acquired in March 2009 
together  with the synthetic spectrum obtained with a sinusoidal 
function fitted to the V band data at a 4.32\,hr 
period.  The latter is shifted in ordinates for clarity.  {\it Bottom:} Folded V (left) and J 
(right) band light curves at this period.}
\label{fig:opticaldft}
\end{figure}

\begin{table}[ht]
\caption{Spectral fit parameters to the broad--band 0.2--100\,keV 
spectrum of XSS\,J1227.}
\label{tab:allspectra}
\begin{center}
\begin{tabular}{lccc}
\hline
\hline\noalign{\smallskip}
Model & Power law & {\sc comptt} & Bremsstrahlung\\
\hline\noalign{\smallskip}
 $N_{\rm H}$  (10$^{21}$cm$^{-2}$)   & 0.98$\pm0.04$  &   0.95$\pm0.05$ & 
0.6$\pm0.03$\\
$\Gamma$  &1.7$\pm0.02$ & --    & --\\
$kT_{\rm seed}$ (keV) & -- & 0.08$\pm0.02$ &--\\
$kT_{\rm plasma}$ (keV) & -- &  53$^{+27}_{-17}$ &--\\
$\tau$  &-- & 1.1$^{+2}_{-0.5}$ & --\\
$kT_{\rm Brems}$ & -- & --  & 12.6$\pm0.6$\\
$\chi^{2}$/d.o.f.  & 976/915  &  982/913  &  1116/914\\
Flux$^{\rm abs,*}_{\rm 0.2-100 keV}$ & 4.2$\pm0.2$   & 3.8$\pm0.5$ & 1.6$\pm0.4$\\
Flux$^{\rm unabs,*}_{\rm 0.2-100 keV}$  & 4.5$\pm0.9$   & 4.1$\pm1.0$ & 1.9$\pm0.9$\\
  &  &  &  \\
\hline\noalign{\smallskip}
\end{tabular}
\end{center}
$^{*}$ in units of 10$^{-11}$erg cm$^{-2}$ s$^{-1}$
\end{table}

\subsection{The optical and near-IR variability}

We analysed the B band \emph{CASLEO}, the V and J band  \emph{REM} light 
curves shown in
Fig.\,\ref{fig:opticalcurves}. All  of them  show large amplitude 
variations ($\sim 40-50\%$) 
on timescales of hours as well as $\sim$20-30$\%$  short term (tens of minutes) variability.
The J band data are  rather poor in statistics and cannot be used to search for 
periodicities. 
 We then Fourier analysed only the B and V band light curves. These do not 
reveal any 
coherent signal at the 
purported period of 859.6\,s, thus confirming the results obtained by 
\cite{Pretorius09}. 
On the other hand,  in the \emph{REM} optical data we  find  significant 
signal at $\nu\sim  6$\,day$^{-1}$ (see Fig.\,\ref{fig:opticaldft}). A sinusoidal fit to the  
\emph{REM} V band light curve covering the three nights 
gives a best fit period of $4.32\pm$0.01\,h.  A comparison of 
the synthetic Fourier spectrum with that of the observed light curve is also shown in 
Fig.\,\ref{fig:opticaldft}. The same period is used to fit the low frequency variability observed 
in the B band \emph{CASLEO} photometry. 
The modulation amplitude is $\sim 50\%$ in 
the B band July 2007 data and is $\sim 40\%$ in the V band March 2009 
observations. 
 The folded light curves in the V and J bands at this period are also shown 
in 
Fig.\,\ref{fig:opticaldft}.  Despite the low statistics of the 
near-IR data, both light 
curves  show a  pronounced minimum at similar phases. A possible secondary mininum seen in the 
 J-band could also be present in the V data, but better data are needed to 
study the near-IR variability. 
The weak quasi-sinusoidal variability detected in the \emph{XMM-Newton} 
quiescent X-ray light 
curve is consistent with this period, suggesting a possible link with the binary orbit in 
the X-ray data.

Furthermore, the B and V band light curves detrended from this periodicity show  large amplitude 
($\sim 40\%$) variability on a timescale of $\sim$ 25-30\,min but this is not found to be 
coherent.
We therefore conclude that this  short term variability is non periodic and hence of flickering 
type.  This is further supported by the photometric variability observed in April 2008 and 
reported  by \cite{Pretorius09}, showing different behaviour from night to night. 
Since the nightly coverage of the present data is at most 4.7\,hr and due 
to the lack of simultaneous X-ray data,  it is difficult to assess  
whether the large amplitude short-term optical variations are linked to the flaring 
activity observed in the X-ray and UV bands. We note however that the UV flares last at least 
1500\,s and hence it cannot be excluded that the optical band is also affected by similar flaring 
behaviour.

\begin{table*}[ht!]
\caption{Spectral fit parameters  of time--resolved spectra  
of  XSS\,J1227 using an absorbed power law.} 

\label{tab:spectra_time}
\begin{center}
\begin{tabular}{lccccc}
\hline
\hline\noalign{\smallskip}
Model & EPIC-pn & EPIC-pn & EPIC-pn & EPIC-pn & EPIC-pn \\
\hline\noalign{\smallskip}
      & Dip    & Out-of-dip  & Flare &  Quiescence & Post-flare dip \\
\hline\noalign{\smallskip}
$\rm N_{H}$  (10$^{21}$cm$^{-2}$)   &1.1$\pm0.3$&  0.97$\pm0.04$ & 
0.78$\pm0.11$ & 0.94$\pm0.04$ &  1.06$\pm$0.06 \\
 $\Gamma$ & 1.7$\pm0.1$  & 1.59$\pm0.02$ &1.59$\pm0.02$  &1.58$\pm0.02$ & 
1.32$\pm$0.3 \\
$\rm N_{H}$  (10$^{22}$cm$^{-2}$) & -- & -- & -- & -- & 6.11$^{+2.4}_{-1.9}$ \\
$C_{F}$ & -- & -- & -- & -- & 0.86$_{-0.11}^{+0106}$ \\
$\chi^{2}$/d.o.f. & 44/55& 932/843 &   245/277  & 915/852  & 25/24 \\
 Flux$^{\rm abs,*}_{\rm 2-10 keV}$ & 0.16$\pm 0.03$ 
& 1.0$\pm0.2$ & 2.6$\pm0.1$  & 0.9$\pm0.2$   & 0.8$\pm$0.1 \\ 
 & & & & & \\
 \hline\noalign{\smallskip}
      & PCA Obs.\,1 & PCA Obs.\,2  & PCA Obs.\,3 &  PCA Obs.\,4 & PCA Obs.\,5\\
\hline\noalign{\smallskip}
   & Quiescence & Flare & Quiescence & Quiescence & Quiescence \\
$\Gamma$ &  1.9 $\pm0.2$     & 1.7$\pm0.1$ & 1.6$\pm0.3$ & 1.7$\pm0.1$ 
&1.7$\pm0.1$\\
$\chi^{2}$/d.o.f. &    21/37 &  27/37 & 20/37 & 35/37   &22/37\\
Flux$^{\rm abs,*}_{\rm 2-10 keV}$ & 1.3$\pm0.2$ & 1.6$\pm0.2$ & 1.2$\pm0.3$ 
&  
1.3$\pm0.2$  & 1.2$\pm0.2$\\
 & & & & & \\
 \hline\noalign{\smallskip}
\end{tabular}
\end{center}
$^{*}$ in units of 10$^{-11}$erg cm$^{-2}$ s$^{-1}$
\end{table*}

\section{The X-ray spectrum of XSS\,J1227}

To investigate the broad-band spectrum and its variability we 
used the \emph{XMM-Newton} EPIC-pn, the \emph{RXTE} PCA and 
\emph{INTEGRAL} ISGRI data with the spectral analysis package {\sc XSPEC} 
version 12.4.

\subsection{The broad-band X-ray spectrum}

The extracted \emph{XMM-Newton} EPIC-pn grand-average spectrum between 
0.2--10\,keV is  featureless with no sign of  an iron complex at 
6--7\,keV. 
The \emph{RXTE} PCA average spetrum extracted between 3--20\,keV also does 
not require any iron line and  does not  show any high energy cut-off.
 
Before combining the different data sets we checked whether the source was 
in a different luminosity state during   the five \emph{RXTE} segments 
and the \emph{XMM-Newton} observation that  were separated by about 
one  year.  Due to the lack of spectral features, we adopted an absorbed 
power law.  The absorption is not needed to fit the \emph{RXTE} PCA 
spectra; we find similar results fixing the absorption to the value 
found from  the \emph{XMM-Newton} data. The results are  reported in  
Table\,\ref{tab:spectra_time}. Both the quiescent flux levels and power 
law indices are about the same in 2007 and 
2009. Also, during both observations the time spent by the source in 
quiescence  is longer than during flares: $\sim 18\%$ during the 
\emph{XMM-Newton}  pointing and $\sim 12\%$ during the \emph{RXTE} 
observation. As shown in Sect.\,4.2, the spectrum does not change 
during the flares. This indicates that the source was in a similar 
luminosity and spectral state at the two epochs. 

We therefore combined the \emph{XMM-Newton} EPIC-pn, 
\emph{RXTE} PCA and  \emph{INTEGRAL} ISGRI data to study 
the broad-band spectrum. We used in our fits  a simple 
absorbed power law, an absorbed cut-off power law and a Bremsstrahlung 
model.  For all the fits we fixed the normalization  constant to the 
\emph{XMM-Newton} data and the normalizations of the PCA and ISGRI data 
were $\sim$1.3. This also indicates that no strong flux variations 
occurred. We find that a power law fits well the data as the inclusion of 
a cut-off is not  statistically significant with a lower limit of 
$55$\,keV. 
A thermal bremsstrahlung model gives a much lower 
fit quality. 
Though  more physically motivated, the Comptonization model 
{\sc comptt}, where soft seed photons with $\rm kT_{seed}$ are 
comptonized by a hot plasma at $\rm kT_{plasma}$ with optical depth 
$\tau$ does not give constrained parameters due to the lack of a 
cut-off at high 
energy in our data.  An additional 
blackbody ({\sc bbody}) is also not 
required by our data. The results of the fits are reported in Table\, 
\ref{tab:allspectra}.

\begin{figure}
\centering
\includegraphics[width=5cm,angle=-90]{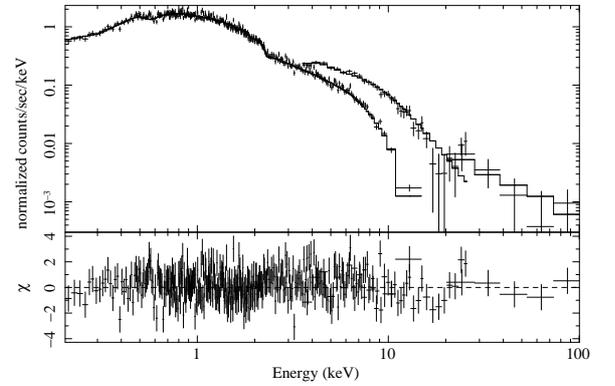}
\caption{The broad-band average X-ray spectrum of XSS\,J1227 using 
the \emph{XMM-Newton} EPIC-pn, the \emph{RXTE} PCA and \emph{INTEGRAL} 
ISGRI spectra, fitted with a simple absorbed power law. The residuals are 
plotted in the lower panel.}
\label{fig:spec_all}
\end{figure}

We therefore conclude that  the broad-band featureless spectrum of 
XSS\,J1227 is well described  by a  weakly absorbed power 
law as shown in  Fig.\,\ref{fig:spec_all}. The 
hydrogen column density 
$\sim 1\times  10^{21}\rm cm^{-2}$ is  compatible with that in the 
direction of the  source ($1.2\times 10^{21}\rm cm^{-2}$) 
\citep{dickeylockman90}, thus indicating an 
origin in the interstellar medium.

\subsection{Time--resolved X-ray spectra}

\begin{figure}
\includegraphics[width=7.cm,angle=-90]{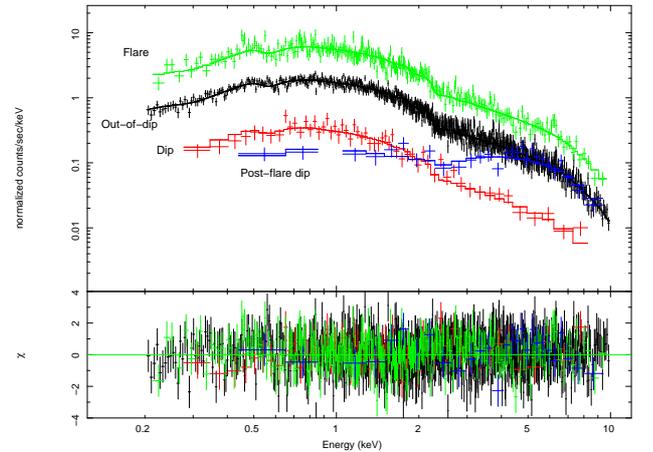}
\caption{Time--resolved \emph{XMM-Newton} EPIC-pn spectra of 
XSS\,J1227 fitted with a simple absorbed power law. 
The post-flare dip spectral fit also includes a partial covering absorber. The lower 
panel shows the residuals}
\label{fig:time_spec}
\end{figure}

The intensity vs hardness ratio diagram shows that the source does not 
significantly change its spectral shape during flares, during the  
quiescent persistent emission and during dips in quiescence. It instead 
changes during the dips observed immediately after the flares. 
We then extracted the \emph{XMM-Newton} EPIC-pn
spectra during the flares (flare spectrum), their associated dips 
(post-flare dip spectrum)  as 
well as during the quiescent dips (dip spectrum), the quiescence 
including (quiescence spectrum) and excluding them (out-of-dip spectrum). 
To fit these data we used an absorbed power law as discussed above.
As expected all spectral fits provide similar power law index within 
errors, except for the post-flare dip spectrum that gives 
$\Gamma=0.22^{+0.13}_{-0.12}$ but $\chi^{2}$/d.o.f. = 49/26. The fit of 
this spectrum  improves {bf by} including a partial covering absorber with 
86$\%$ 
covering fraction and $\rm N_{H} = 6.1\times 10^{22}\,cm^{-2}$. 
With this component the power law index increases to $\Gamma$=1.3 and is 
consistent within errors with that found in the other fits. 
Furthermore, the flare spectrum  shows hints of an emission feature
at the iron complex. However, the inclusion of a gaussian line, 
found at 6.2\,keV, gives a slight though not 
significant, improvement 
to the fit. 
The results are reported in  Table\,\ref{tab:spectra_time} and shown 
in Fig.\,\ref{fig:time_spec}.

\section{The combined X-ray and gamma ray spectrum }

\begin{figure}
\includegraphics[width=6.cm,angle=-90]{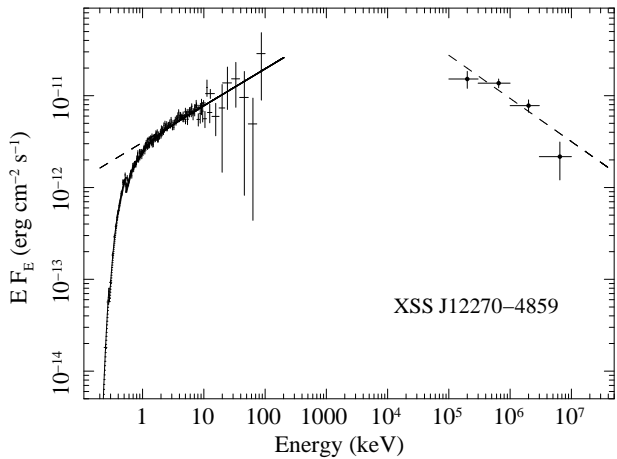}
\caption{The broad band X-ray to high energy gamma ray spectrum combining 
the \emph{XMM-Newton} EPIC-pn, the \emph{RXTE} PCA and \emph{INTEGRAL} 
ISGRI rebinned data and the \emph{Fermi} LAT dat together with their 
respective best fit power law models.  Solid line and dashed lines 
represent the absorbed and  unabsorbed best fit spectral models.}
\label{fig:flux_x_gamma}
\end{figure}

 Considering the possible association with 1FGL\,J1227.9-4852, we have 
combined the fluxes in the four \emph{Fermi} LAT bands 
 and those in the \emph{XMM-Newton} EPIC-pn, 
\emph{RXTE} PCA and   \emph{INTEGRAL}  ISGRI bands. The energy spectrum is 
shown in  Fig.\,\ref{fig:flux_x_gamma} together with the corresponding best 
fit models
with power law indexes 1.70 (low energy, Table\,\ref{tab:allspectra}) and 
2.45 (high energy, sect.2).
The shape of the combined spectrum suggests that if the low and high 
energy emissions are related, the peak energy should be between 
1-100\,MeV. 
A similar spectrum is observed in LS\,I+61$^o$303; \citep{Chernyakova06}, 
though it could be modeled under the assumption of a pulsar powered 
source. Lacking of knowledge on the nature of the source, it is not 
possible to use physical models at this stage. With our measured flux 
values, the 
0.1-100\,GeV/0.2-100\,keV luminosity ratio is of the order of $\sim$0.8. 
Hence, if the identification is correct, the GeV emission is a significant 
component of the total energy output.

\section{Discussion}

We have presented  X-ray, UV and optical/nIR observations of 
XSS\,J1227.  We also found  strong indication that this source has a 
high energy GeV counterpart as detected by the \emph{Fermi} 
satellite.

\noindent   XSS\,J1227 shows remarkable large amplitude variability 
from X-rays to optical/near-IR. However, we did not detect the claimed 
859.6\,s periodicity in any data set from  
X-rays to  UV/optical and near-IR ranges. 
The \emph{XMM-Newton} X-ray light curve is characterized by short   
aperiodic variations consisting of flares and dips. The latter are 
observed during quiescence as well as immediately after the 
flares. This peculiar behaviour is also detected in a \emph{Suzaku} 
observation  \citep{Saitou} carried out five months before the 
\emph{XMM-Newton} pointing. 
This  is not reported to be present  in the 
  \emph{RXTE} observations performed in 2007 by \cite{Butters08}, but a 
re-analysis of the same data reveals instead a similar behaviour 
as detected by \emph{XMM-Newton} and \emph{Suzaku}. Also, the purported 
859.6\,s period is not detected in the same \emph{RXTE} data.  
The broad-band X-ray  spectrum is essentailly featureless  and 
 is well described by an absorbed 
simple power law with $\Gamma\sim 1.6$. An emission feature at 6.2\,keV 
could be present during flares.  
From both temporal and 
spectral  characteristics, we therefore conclude that  XSS\,J1227 is a 
persistent highly variable hard X-ray  source  that does  not 
share any of the typical X-ray characteristics of  magnetic CVs,  
especially  of the IP type, and any commonality  
of CV flares as for instance seen in AE\,Aqr \citep{Choi} or  UZ\,For 
\citep{Still} and AM\,Her
\citep{deMartino}. A similar conclusion was drawn 
by \cite{Saitou}.

\noindent The evolution of the X-ray events is rather 
similar consisting of flare-dip pairs where the duration and intensity of 
flare and associated dips appear correlated. While no spectral changes are
observed during flares with respect to quiescence, the spectrum hardens 
during the post-flare dips. A dense ($\rm N_H = 6.1\times 
10^{22}\,cm^{-2}$) absorbing material covering about $\sim 86\%$
of the X-ray  source is required to fit these dips.
This is suggestive of a flow of cool material appearing after the flares.
These flares also occur  in the UV but with longer duration. 
The UV variations lag by more than 300\,s the X-ray ones. The UV flux
gets bluer during the flares  than in quiescence. It is then possible that 
large amplitude, 
long  term variations  first affect the outermost parts of an accretion 
disc  that are cooler and  then propagate towards smaller radii. The UV decay is 
delayed suggesting that the UV is also affected by reprocessing of X-rays 
after the flare.

\noindent Pronounced aperiodic X-ray dips are 
observed when the source is in quiescence with no significant spectral 
changes. On the other hand, no 
dips are  detected in the 
UV  band suggesting that they originate from random  occultations by 
material very close to the X-ray source. 

\noindent New optical and near-IR photometry reveals large amplitude 
(up to $50\%$) variability.  A  periodicity at 4.32\,hr is derived from 
the optical data.  The modulation at this period
is single peaked in the optical whilst it is double-humped in the near-IR 
band. A marginal evidence of a low ($4\%$) amplitude variability at this period is also 
found in the X-ray and UV ranges. 
If this period is linked to the orbital binary period it implies 
that XSS\,J1227 is a Low Mass X-ray Binary (LMXRB). The near-IR 
double humped modulation could then be due to ellipsoidal variations from 
the non-spherical low-mass donor star. The amplitude is  determined 
by the orbital inclination angle $i$ of the binary \citep{Gelino}.
If it is indeed the case  the binary inclination: $i\ga 60^o$. On the 
other hand 
eclipses are not observed suggesting $i<75^o$.

\noindent We also revised the optical spectrum presented in 
\cite{Masetti06} and 
confirm the equivalent 
 widths of major Balmer emission lines. We note however that due to the 
low spectral resolution, the He\,II (4686$\AA$) line 
is  blended with  C\,III (4650). We then  measure E.W.(He\,II)=4$\pm 1 
\AA$, thus being much weaker than previously measured.
Hence, the H$_{\beta}$ and He\,II E.W. ratio, when compared to that of CVs and 
LMXRBs  \citep{vanparadis84} locates XSS\,J1227 between the two 
object class  locii. Furthermore
 the X-ray to optical flux ratio ranges between 48 (flares) and 17 
 (quiescence). This value is   larger than that of CVs and magnetic 
systems 
and lies in the low value range of LMXRBs.

\noindent If XSS\,J1227 has an orbital period of 4.3\,hr the donor 
is expected to be a low mass star with $\rm M_2 \sim$ 0.3-0.4\,M$_{\odot}$ 
\citep{Smith_Dhillon98,Pfahl03,knigge06} of spectral 
type between $\sim$M3.1-M3.3 and 
with an absolute  near-IR J band 
magnitude $\rm M_J \sim$ 6.7\,mag \citep{knigge06}. The faintest 
measured  J-band magnitude of XSS\,J1227 (16.9\,mag), when corrected 
for interstellar absorption, A$_J$=0.12\,mag, obtained using the 
derived hydrogen column density from X-ray spectra, would imply a 
distance d $\ga$ 1.1\,kpc, if the near-IR emission is totally due to the 
secondary star.  This minimum distance could be consistent with 
the presence of near-IR ellipsoidal variations. Also, the source is 
located at $\rm \sim 13^o$ in galactic latitude and, if it is in the 
galactic disc, its distance should not be exceedingly large. 
The X-ray bolometric luminosity is then $\rm L_X \ga 6\times 10^{33}\,\rm erg\,s^{-1}$
suggesting a LMXRB accreting at a low rate. 

 The present analysis therefore favours XSS\,J1227 as a 
peculiar, low-luminosity LMXRB. Its flaring characteristics, consisting of
flare-dip pairs are reminiscent of the type II bursts observed in 
the bursting pulsar GRO\,J1744-28  \citep{Nishiuchi} or in the Rapid Burster \citep{lewin}. However in these 
sources the energetics and timescales are much  different as well as the 
spectral dependence of their associated dips. GRO\,J1744-28 could be more 
similar to  XSS\,J1227, as the bursts (giant and small) 
do not show significant changes in spectral shape and show a good 
correlation between burst fluence and flux deficiency in the associated 
dips \citep{Nishiuchi}. However during the post-flare dips, 
GRO\,J1744-28 shows no 
 spectral changes and the burst fluence is related to the time when
the source is in the persistent quiescent state. This is not the
case of XSS\,J1227. The Rapid Burster, instead shows 
different  spectral behaviour during the post-flare dips as well as the 
pre-flare ones \citep{lewin}. Hence, XSS\,J1227 could share common 
properties  with type II bursts of the above sources, but with some 
differences. 

\noindent Type II bursts are believed to be due 
to instabilities in the accretion disc that  produce a rapid accretion 
onto 
the compact object depleting a reservoir in the inner disc regions. This is 
 replenished immediately after the burst, thus producing a flux depression
that does not affect the X-ray spectrum. However, type II bursts have 
not always the same morphology. For istance SMC X-1, a high 
mass X-ray binary, does not show post-flare dips
\citep{Angelini,Moon03}. Given the few sources known so far we cannot 
exclude that XSS\,J1227 is a type II low-level bursting source.
Its UV activity starts before that in the X-rays. Also, the timescale of 
UV and X-ray flares  is longer than that
observed in the type II bursters and of the order of the  
free fall  time $\rm t_{ff} \ga 603\, (M_1/M_{\odot})^{-1/2}\,s$  
from the Roche-lobe radius of the compact object ($\rm 
R_{L,1} \ga 0.66\,R_{\odot}$ for $\rm q=M_2/M_1 \leq 0.8$, for a $\rm 
P_{orb}$=4.32\,hr and
 M$_2\ga 0.3-0.4\,M{\odot}$).   The further presence of a 
partial (almost total) covering absorber during 
the post-flare dips could be the result of 
the replenishing of a larger portion of the
accretion disc. This could be corroborated by the relatively long ($\sim$300--600\,s) 
post-flare dips. 

\noindent While dips during quiescence are observed in many LMXRBs 
such as EXO\,0748-676 \citep{Bonnet-Bidaud01} or 4U\,1916-05 
\citep{Callanan93}, these occur at specific orbital phases and are 
accompained by a    hardening of the source  due to absorption of matter 
from the outer rim of  the disc.  XSS\,J1227 is hence different 
from the LMXRB  dippers.

\noindent Type II bursters and LMXRB  dippers are known to harbour a 
neutron star 
being either pulsars (SMC X-1 
and GRO\,J1744)  or also showing type\,I bursts, as the Rapid Burster,  
that are   a signature  of thermonuclear flashes  on a 
neutron star.  
With  the present 
 data it is not possible to establish whether the compact object in 
XSS\,J1227 is a pulsar.

 The \emph{Fermi}  detection in the GeV range of the 
source 1FGL\,J1227.9-4852, consistent with the XSS\,J1227 position, may 
further strengthen the interpretation of a LMXRB. Worthnoticing is that 
only a few X-ray binaries are detected so far with \emph{Fermi}. In the 
first release of the  \emph{Fermi} source catalogue, only three are 
classified as High Mass X-ray Binaries (HMXRB) and five as LMXRBs, but 
among them three are associated with globular clusters, the other two are 
identified with the Galactic centre and the supernova remnant G332.4-00.4. 
The sources unambiguously identified by their periodicities are the HMXBs, 
1FGL\,J0240.5+6116 (LS\,I+61$^o$303; \citep{abdo09a}), 1FGL\,J1826.2-1450 
(LS\,5039; \citep{abdo09b}) and 1FGL\,J2032.4+4057 (Cyg\,X-3; 
\citep{abdo09c}). LS\,I+61$^o$303 ($\rm P_{orb}$=26.5\,d) and LS\,5039 
($\rm P_{orb}$=3.9\,d) are long period systems for which the high energy 
gamma ray emission dominates with rather similar 
(0.1--100\,GeV)/(0.2--100\,keV)
luminosity ratio of $\sim$6.8 and 
$\sim$6.2, respectively; while for the shorter period system Cyg\,X-3 
(($\rm 
P_{orb}$=0.2\,d) this ratio is at much lower value of $\sim$ 
0.01-0.03~\footnote{ The  broad-band 0.2--100\,keV fluxes of 
LS\,I+61$^o$303, LS\,5039 and Cyg\,X-3 are taken from 
\cite{Chernyakova06}, \cite{Takahashi09} and 
\cite{Hjalmarsdotter09}, respectively.}.
With a value of $\sim$0.8, XSS\,J1227 could therefore be an intermediate 
system between these two regimes. 

\noindent It is also possible that 1FGL\,J1227.9-4852 is a separate 
confusing source, like a Geminga-like pulsar. This possibility should not 
be discarded until a detailed temporal analysis of the GeV 
emission is  performed. This will allow  to infer whether a flaring-type 
activity on  similar  timescale as  that observed in the X-rays  and/or a 
periodic variability, either at the putative orbital period or neutron star 
spin,  can be detected.

\noindent We also searched for a radio counterpart in the RADIO (Master 
Radio) catalogue  available  at HEASARC  archive 
~\footnote{ http://www.heasarc.gsfc.nasa.gov/W3Browse/all/radio.html}.  
The only radio source within the 6' \emph{Fermi} LAT error radius  
is catalogued in the Sydney University Molonglo Sky  Survey, 
SUMSS\,J122820-485537 with a 843\,MHz flux of 82.2$\pm$2.7\,mJy. This 
source, also shown in Fig.\,\ref{fig:lat_image}, is found  at 
5.22'  from 
1FGL\,J1227.9-4852 and at 4.12' from the 
XSS\,J1227 optical position. The radio positional accuracy is quite 
high with an ellipse uncertainty semi-major axis of 4.3''  and 
semi-minor  axis of 4.0''. Hence, although within the  \emph{Fermi} LAT 
95$\%$ confidence region, the association to XSS\,J1227 is quite 
unsecure.
   
The data presented here have therefore shown that  XSS\,J1227 is a 
rather 
atypical LMXRB, that might  reveal a new class of  low-luminosity X-ray 
binaries or a peculiar  accretion regime. 
 The possible association with the GeV 
\emph{Fermi} LAT source 1FGL\,J1227.9-4852 further suggests 
such peculiarity. To shed light into its intriguing nature 
a timing analysis  of the high energy emission is essential for 
a secure identification with  XSS\,J1227. Also,  a long-term X-ray 
monitoring to constrain the 
 flaring and dipping behaviour and to infer whether this source  undergoes 
higher states or bursts  is needed as well as
 time--resolved spectroscopy of 
the optical  counterpart  to confirm whether the photometric period is 
the binary orbital period. 

\begin{acknowledgements}
DdM,  TB and NM acknowledge financial support from ASI under contract 
ASI/INAF I/023/05/06 and ASI/INAF 
I/088/06/0 and  also from INAF under contract PRIN-INAF 2007 
N.17.
 We gratefully acknowledge the help of  E. Bonning in the extraction 
of  counts map from the  \emph{Fermi} LAT archive.
\end{acknowledgements}

\bibliographystyle{aa}
\bibliography{AA1380}
\end{document}